\documentclass[aps,prl,twocolumn,floatfix,10pt,amssymb,amsfont,amsmath,superscriptaddress,float,longbibliography]{revtex4-2}
\usepackage{graphicx}
\usepackage{dcolumn}
\usepackage{bm}
\usepackage{color}
\usepackage[normalem]{ulem} 

\usepackage{hyperref}
\begin{document}

\preprint{APS/123-QED}

\title{Emergent Andreev Reflection from a Lattice Duality Defect}

\author{Atsushi Ueda}
\thanks{These two authors contributed equally.}
\affiliation{Department of Physics and Astronomy, Ghent University, Krijgslaan 281, 9000 Gent, Belgium}
\author{Tokiro Numasawa}
\thanks{These two authors contributed equally.}
\affiliation{Institute for Solid State Physics, The University of Tokyo, Chiba, 277-8581, Japan}
\author{Boris De Vos}
\affiliation{Department of Physics and Astronomy, Ghent University, Krijgslaan 281, 9000 Gent, Belgium}
\author{Masataka Watanabe}
\affiliation{Faculty of Science, The University of Tokyo, Tokyo 113-0033, Japan}


\date{\today}

\begin{abstract}
Andreev reflection converts an incoming fermion into an outgoing hole and is usually tied to a superconducting interface. We show that an analogous charge-conjugating boundary condition emerges from a purely lattice duality defect. Starting from a Majorana representation of the transverse-field Ising chain, we construct a folded lattice model in which a boundary Majorana impurity implements a one-site translation of a staggered Majorana chain. In the continuum, this translation becomes a chiral fermion-parity defect: it flips the sign of the only left-moving Majorana mode while leaving the right-moving mode unchanged. When the two Majorana modes are recombined into a complex fermion in the folded geometry, this sign flip becomes the Andreev-like boundary condition. 
Our lattice formulation gives a microscopic interpretation of the Emery--Kivelson boundary of the two-channel Kondo problem and of Maldacena--Ludwig monopole scattering, while identifying the boundary as the interface between a Kitaev-chain SPT phase and a gapless chain.
The same Majorana translation defect also provides a lattice realization of an axial $U(1)_A$-symmetric charge-flip boundary.
\end{abstract}

\maketitle

\paragraph{Introduction.}
What comes back when an electron is scattered? In ordinary systems, the answer seems obvious: an electron. But in strongly correlated quantum systems, even this simple expectation can fail. 
A famous example is the Andreev reflection~\cite{andreev1965thermal}. At the interface between a normal metal and a superconductor, an incoming electron does not bounce back as an electron. It is reflected as a hole. The reason is that, below the superconducting gap, a single electron cannot enter the superconductor on its own. It must recruit another electron from the normal metal, and the two electrons enter as a Cooper pair. What is left behind is precisely a hole: 
\begin{equation*}
    \psi_{\mathrm{out}} \sim \psi^\dagger_{\mathrm{in}}.\label{eq:Andreev}
\end{equation*}

Interestingly, the Andreev reflection reappears in several seemingly unrelated areas of physics. Emery and Kivelson showed that it arises in the low-energy effective field theory of the two-channel Kondo impurity problem~\cite{EK_1992}. Maldacena and Ludwig later pointed out that a similar field theory describes the scattering of chiral fermions from a magnetic monopole~\cite{Maldacena_1997}. Of course, neither of these examples involves an actual superconductor. What, then, is the universal mechanism behind this Andreev-like scattering? 

This question is particularly sharp on the lattice. The relevant field theories involve chiral fermions, whose lattice realization is famously constrained by the Nielsen-Ninomiya no-go theorem~\cite{Nielsen:1980rz}. Thus, even formulating a simple lattice picture of these scattering processes is already a nontrivial problem.

In this Letter, we address this question directly on the lattice. We show that these scattering boundaries are topological duality defects in disguise, with a Majorana zero mode sitting at their core. In this language, the mechanism becomes remarkably simple: the Majorana zero mode shifts the Majorana operators across the interface, and this translation appears in the continuum as emergent Andreev reflection. This picture arises naturally at an interface between the symmetry-protected topological (SPT) phase of the Kitaev chain~\cite{Kitaev:2000nmw} and a gapless free-fermion chain.
We further show in the Supplemental Material that, although the boundary flips the ordinary vector charge, it preserves an axial $U(1)_A$ symmetry on the lattice. 
This gives a lattice realization of an axial symmetry-preserving boundary.

\paragraph{Duality defect and lattice Emery--Kivelson model.}
Let us start with a familiar example: the transverse-field Ising model. This is an exactly solvable and canonical model of a quantum ferromagnet~\cite{Pfeuty_1970}. We consider the slightly modified open-chain
Hamiltonian
\begin{align}
    H_{\mathrm{Ising}} = -\sum_{j=1}^{N-1} X_{j} X_{j+1} -g\sum_{j=1}^{N-1} Z_j, \label{eq:Ising}
\end{align}
where $X$ and $Z$ denote Pauli matrices.
A sharp-eyed reader may notice something fishy: the term $Z_N$ is missing. This is not an oversight, but an intentional omission. The last spin is still coupled to the rest of the chain through the term $X_{N-1}X_{N}$, and is therefore not simply an independent spin. Nevertheless, direct diagonalization shows that Eq.~\eqref{eq:Ising} has a two-fold ground state degeneracy for any value of $g$. One may now try to make the model even more suspicious by also removing the transverse field on the first site, $Z_1$. Surprisingly, the ground state degeneracy remains two-fold. It is as if no additional degree of freedom has been added this time. This is the small puzzle from which we begin.

\begin{figure}[bt]
    \centering
    \includegraphics[width=\linewidth]{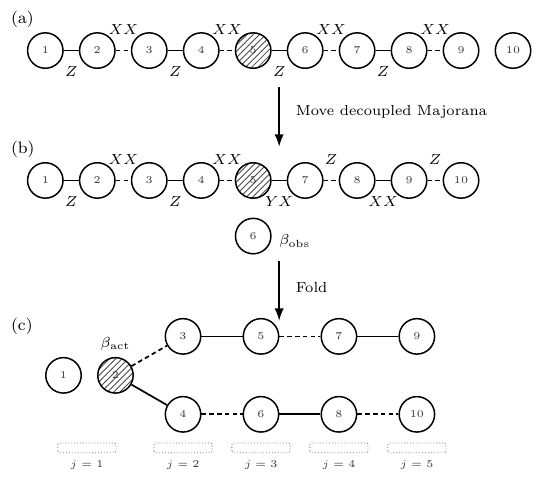}
    \caption{Three representations of the same physical theory. A pair of adjacent Majorana fermions defines a local Hilbert space. 
(a) The original transverse-field Ising chain. Solid and dashed lines denote couplings of alternating strength. 
(b) The same model after sliding the decoupled Majorana $\beta_{\mathrm{obs}}$. $\beta_{\mathrm{obs}}$ forms a pair with the fifth Majorana (which we call the active Majorana, shaded in grey) for spanning the Hilbert space.
(c) The folded representation of the impurity Ising chain.}
    \label{fig:folded-model}
\end{figure}

The puzzle becomes delightfully transparent after rewriting the model in terms of Majorana fermions. Using the Jordan-Wigner transformation \cite{Jordan_Wigner_1928}
$\chi_{2j-1} = \left(\prod_{k=1}^{j-1} Z_k\right)X_j,\, \chi_{2j} = \left(\prod_{k=1}^{j-1} Z_k\right)Y_j.$ 
Then the Ising Hamiltonian becomes
\begin{align}
    H_{\mathrm{Majorana}} = i\sum_{j=1}^{N-1} \left[g\chi_{2j-1}\chi_{2j} +\, \chi_{2j}\chi_{2j+1}\right].\label{eq:majorana}
\end{align}
This structure is illustrated in Fig.~\ref{fig:folded-model} (a) for $N=5$. The missing $Z_N$ term leaves the last Majorana $\chi_{2N}$ completely decoupled. This is an exact Majorana zero mode. But a single Majorana cannot by itself form a qubit; a physical fermion requires a pair of Majoranas. The partner of $\chi_{2N}$ is an extensive mode in the bulk~\footnote{it is the extended zero mode 
$\gamma = \frac{1}{\sqrt{\mathcal{N}}}\sum_{k=1}^{N} g^{k-1}\chi_{2k-1},$
where $\mathcal{N}$ is a normalization factor.}. The two Majorana zero modes $\gamma$ and $\chi_{2N}$ combine into one free fermionic mode, giving the observed two-fold degeneracy.

Now the second observation is no longer a mystery. If we also remove the first transverse field $Z_1$, the first Majorana $\chi_1$ becomes explicitly decoupled. In this modified model, the extended zero mode $\gamma$ is simply replaced by the boundary Majorana $\chi_1$. Thus, the two zero modes are now manifestly $\chi_1$ and $\chi_{2N}$, but they still form only one fermionic degree of freedom. Removing $Z_1$ therefore does not create an additional qubit: it makes the hidden partner of $\chi_{2N}$ visible. This resolves the puzzle. This phenomenon is analogous to the boundary degeneracy of an SPT phase with open boundary conditions \cite{Kennedy_1990}.
The important idea here is that the Majorana zero mode is topological: since it is decoupled from the Hamiltonian, it can be moved throughout the system without changing the energy spectrum. Let us move it, for example, to the middle of the chain $j=I$ as shown in Fig.~\ref{fig:folded-model} (b). Once again, this does not change the physics. However, the appearance of the Hamiltonian changes as~\cite{Ueda:2025ecm,Aasen:2016dop,Antinucci:2026}
\begin{align}
    H_{\mathrm{impurity}} = &-\sum_{j=1}^{I-2}X_jX_{j+1} -g\sum_{j=1}^{I-1} Z_j \nonumber \\
     &-X_{I-1}X_{I}-gY_I X_{I+1}\nonumber \\
            &-\sum_{j=I+1}^N Z_j -g\sum_{j=I+1}^{N-1} X_jX_{j+1}.\label{eq:H_impurity}
\end{align}
The three lines of Eq.~\eqref{eq:H_impurity} naturally correspond to the original Ising chain, the impurity coupling, and the dual Ising chain, respectively. The right side of the Hamiltonian has been transformed into the Kramers--Wannier (KW) \cite{Kramers_Wannier_1941} dual Hamiltonian by sliding the free Majorana to the left. At first sight, this looks like magic. But it is not: in the Majorana language, KW duality is simply a one-site translation \cite{Seiberg:2023cdc},
$$\chi_{j}\rightarrow\chi_{j+1}.$$
Thus, when the free Majorana slides to the left, it pushes the remaining Majoranas on its right by one site. This is the microscopic origin of the duality transformation. 

The next step is even more revealing. We fold this impurity chain at its center, as shown in Fig.~\ref{fig:folded-model} (c). The new Hamiltonian is 
\begin{align}
    H_{EK} =& i\sum_{\,j \, \mathrm{even}}^{N-1}\left(g\chi^1_{j+1}\chi^1_{j} + \chi^2_{j}\chi^2_{j+1} \right) \nonumber \\
    &+i\sum_{j > 1\, \mathrm{odd}}^{N-1}\left(\chi^1_{j+1}\chi^1_{j} + g\chi^2_{j}\chi^2_{j+1} \right) \nonumber \\
    &+i\chi^1_2\beta_\mathrm{act} + ig\beta_\mathrm{act}\chi^2_2,
\end{align}
where $\beta_\mathrm{act}$ is the active Majorana mode serving as the boundary in the folded model illustrated in Fig.~\ref{fig:folded-model} (c) with a shaded circle, and the superscripts distinguish between the top and bottom Majoranas. After folding, the qubit Hilbert space is spanned by different pairings of Majorana operators, namely vertical pairs $\chi^1_{j}$ and $\chi^2_{j}$. \textcolor{black}{In this pairing, the Majorana operators are ordered as
$\chi^1_j = \chi_{2j-1},\, \chi_j^2 = \chi_{2j}.$}
After the particle-hole transformation \footnote{This is a unitary transformation. In the Ising impurity model, this is a rotation of the spin basis for the half-chain.},
$$\chi^2_{2j}\rightarrow -\chi_{2j}^2,$$
the spin Hamiltonian becomes 
\begin{align}\label{eq:EKmodel}
    H_{\mathrm{EK}} =& \sum_{j \, \mathrm{even}}^{N-1} (X_{j} Y_{j+1}- gY_jX_{j+1})  \nonumber \\
    &+\sum_{j > 1\, \mathrm{odd}}^{N-1} (gX_{j} Y_{j+1}- Y_jX_{j+1}) \nonumber \\
    &+ X_1(X_2+g\,Y_2).
\end{align}
Here, EK stands for Emery--Kivelson. \textcolor{black}{The details of this transformation are in the Supplemental Material}. After refermionization, this model takes the form
\begin{align}
    H_{\rm EK}
&=
i\beta_{\mathrm{act}}
\left[
c_2+c_2^\dagger
+ig(c_2-c_2^\dagger)
\right]\nonumber\\
&+
i(1+g)\sum_{j=2}^{N-1}
(c_j^\dagger c_{j+1}-c_{j+1}^\dagger c_j )\nonumber\\
&+
i(1-g)\sum_{j=2}^{N-1}(-1)^j
(c_j^\dagger c_{j+1}^\dagger-c_{j+1}c_j).\label{eq:EK_complex_fermion}
\end{align}

Thus, this model is a complex fermionic chain coupled to a single Majorana fermion at the boundary. Since this structure mirrors the field-theoretical description of the Emery--Kivelson boundary, we call it the lattice Emery--Kivelson model. The remaining question is whether this model realizes the Andreev reflection as predicted by the corresponding continuum theories~\cite{EK_1992,PhysRevLett.103.237001}.

\paragraph{Scattering dynamics.}
\begin{figure}[tb]
    \centering
    \includegraphics[width=\linewidth]{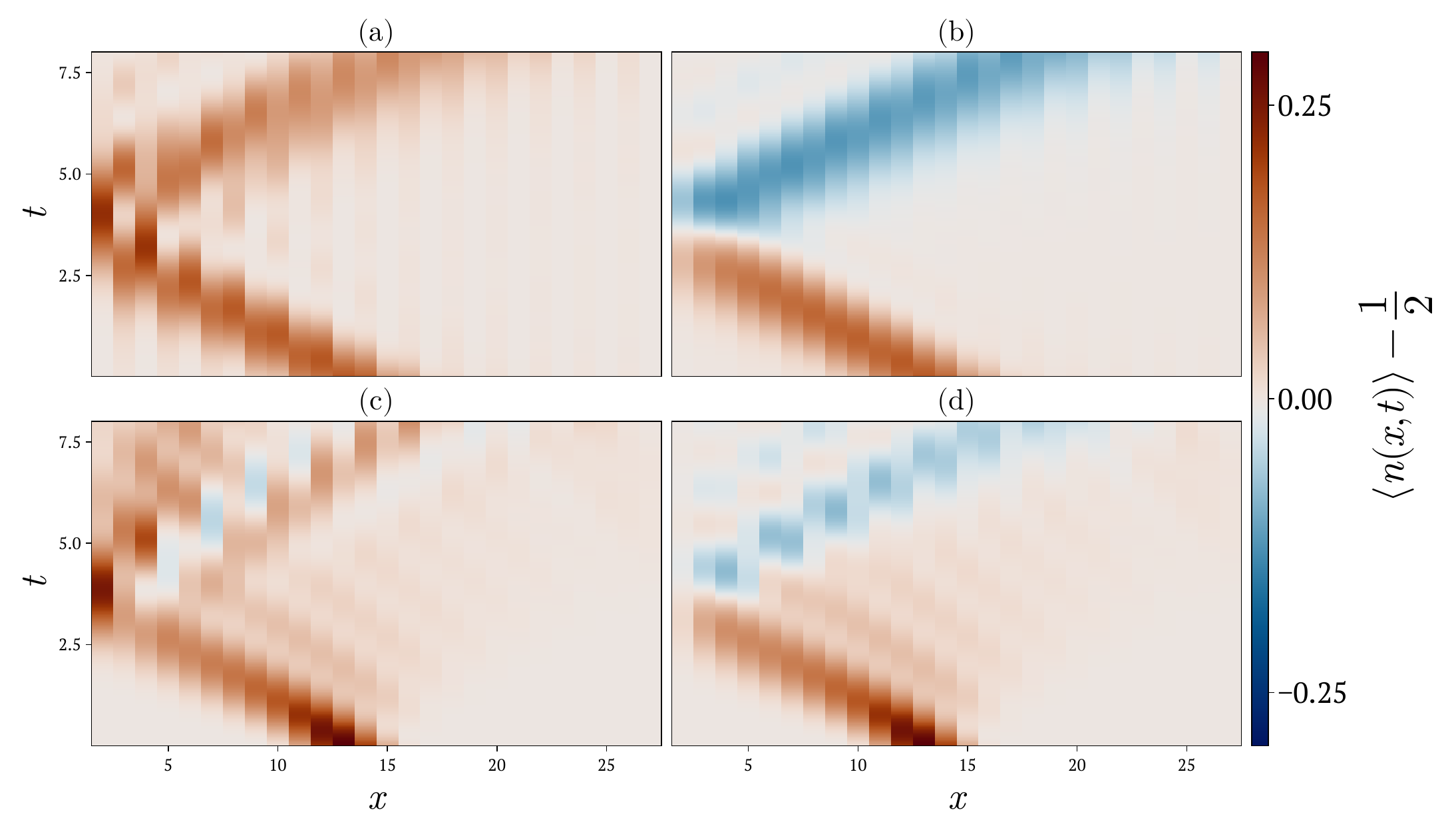}
    \caption{Scattering dynamics for the free boundary condition in (a,c) and the Emery--Kivelson (EK) boundary condition in (b,d), at $g=1$ in (a,b) and $g=3$ in (c,d). The wave packet is created by applying a fermion creation operator to the ground state which propagates toward the boundary at $x=1$, where it is reflected. For the free boundary condition in (a,c), the charge of the excitation is preserved upon reflection. For the EK boundary condition in (b,d), the charge is flipped.
}
    \label{fig:numerics}
\end{figure}
The answer is yes. This folded Ising impurity model indeed realizes the charge-flip boundary. To demonstrate this, we numerically simulate the scattering dynamics using matrix product states (MPS) \cite{PhysRevLett.69.2863,PhysRevB.48.10345,Schollwock_2011,Verstraete:2008cex} and the time-dependent variational principle (TDVP) algorithm~\cite{PhysRevLett.107.070601}. We first obtain the many-body ground state using the density matrix renormalization group (DMRG) algorithm and create a Gaussian wavepacket $\hat{W}({x_0},k) = \sum_x e^{-(x-x_0)^2/\sigma^2}e^{ikx}c^\dagger_x$ on top of it as $|\psi_{\mathrm{init}}\rangle = \hat{W}|\psi_{\mathrm{GS}}\rangle$. Since the ground state is at half filling, the scattering process can be monitored through the charge fluctuation $\delta n(x) = n(x) -\frac{1}{2}$. The results are shown in Fig.~\ref{fig:numerics}. \textcolor{black}{We time-evolved the wavepacket long enough to observe the scattering process at the boundary of interest to the left. The other boundary is set to be free for simplicity.} For the free boundary condition, which is Eq.~\eqref{eq:EK_complex_fermion} without the first term, the incoming electron is reflected as an electron, as expected, shown in Fig.~\ref{fig:numerics} (a,c). By contrast, for the EK boundary condition, Fig.~\ref{fig:numerics} (b,d), the incoming electron is reflected as a hole. The lattice model therefore realizes precisely the desired Andreev reflection. 

This leaves us with the central question: we have only folded the Ising chain at a duality defect. Why does Andreev reflection emerge? 
\begin{figure}[t]
    \centering
    \includegraphics[width=\linewidth]{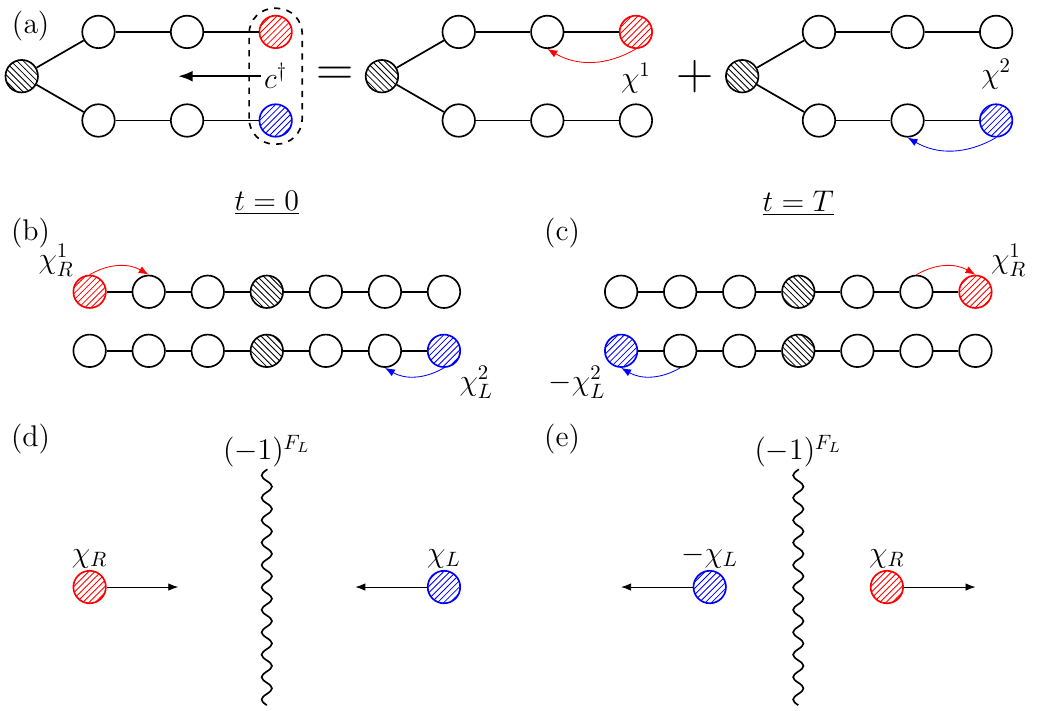}
    \caption{Andreev reflection in the lattice EK model.
(a) Before unfolding, a wavepacket of the complex fermion approaching the boundary decomposes into a pair of chiral Majorana fermions hopping in the same direction.
(b) After unfolding, the pair of chiral Majorana fermions move in opposite directions towards the impurity.
(c) The boundary Majorana impurity implements a one-site translation, producing a minus sign only for the left-moving mode.
(d,e) Continuum counterpart of the same process. The Majorana impurity acts as a chiral fermion-parity defect.
}
    \label{fig:low-energy}
\end{figure}
\paragraph{Charge flip is one-site translation.}
The answer is surprisingly simple: it is, once again, a one-site translation. To see this, imagine creating a wavepacket at site $j$ of Fig.~\ref{fig:folded-model} propagating toward the impurity. The fermion creation operator is built from two Majorana operators,
$$c_j^\dagger = \frac{1}{2}(\chi^1_j - i\chi^2_j),$$
where the superscripts 1 and 2 label the top and bottom chains. In the folded geometry, the wavepacket simply moves toward the boundary, see Fig.~\ref{fig:low-energy} (a). After unfolding, however, the same process is better viewed as a pair of Majorana excitations moving in opposite directions, as shown in Fig.~\ref{fig:low-energy} (b,c). Schematically, the initial state is 
$$\psi_{\mathrm{in}} \sim \chi_R^1 - i \chi^2_L.$$ 
Under time evolution, these two Majorana excitations pass one another and exchange their positions. Without the Majorana impurity, nothing dramatic happens: this is just a free propagation, and the outgoing state is the same as the incoming one. 

So what does the EK boundary do? In the unfolded picture, the active Majorana mode $\beta_{\mathrm{act}}$ sits at the center and shifts the Majorana operators on one side by one lattice site. This is precisely the one-site translation discussed above. The only trick is that this innocent-looking translation acts differently on the two low-energy chiral modes. At $g=1$, the bulk Hamiltonian~\eqref{eq:EK_complex_fermion} has the energy dispersion $E(k) = 2\sin(k)$. The low-energy modes live near $k=0$ and $\pi$, so the lattice fermion operator decomposes as
$$c_n \sim \psi_R(x) + (-1)^n \psi_L(x).$$
A one-site translation therefore leaves the right mover unchanged, but gives the left mover a minus sign. Returning to Fig.~\ref{fig:low-energy}, the blue Majorana excitation picks up this minus sign as it crosses $\beta_{\mathrm{act}}$, as shown in Fig.~\ref{fig:low-energy} (c). The final state is then
$$\psi_{\mathrm{out}} \sim \chi_R^1 + i \chi^2_L = \psi_{\mathrm{in}}^\dagger.$$ 
In other words, the one-site Majorana translation disguises itself in the continuum as Andreev reflection. 

This also dovetails with the continuum field theory picture. 
The continuum theory is described by a free Dirac fermion on the half-line,
\begin{align}
S &= \int_{-\infty}^{\infty} dt \int_0^{\infty} dx
\big (i \bar{\psi}\gamma^{\mu}\partial_{\mu}\psi  \notag \\
& \quad  - m \bar{\psi}\psi 
- \frac{\Delta}{2} (\bar{\psi} \psi^c + \bar{\psi}^c \psi)\big).
\end{align}
Here $\psi = (\psi_L, \psi_R )^T$ is a two-component field, whose components are the left-moving mode $\psi_L$ and the right-moving mode $\psi_R$.
$\psi^c $ denotes the charge-conjugate spinor.
In our basis, charge conjugation is represented by complex conjugation $\psi^c = \psi^*$.
$\gamma^{t} = i \sigma^y$ and $\gamma^x = \sigma^x $ are gamma matrices in $1 + 1$ dimensions and $\bar{\psi} = \psi^{\dagger}\gamma^t$.
At the boundary $x=0$, we impose the boundary condition
\begin{align}
\psi_L(t,0) = \psi_R^{\dagger}(t,0). \label{eq:CCbdy}
\end{align}
In $1+1$ dimensions, the Dirac fermion can be decomposed into two Majorana fermions as
\begin{align}
\psi = \lambda^1 + i \lambda^2 =
\begin{pmatrix}
\lambda_L^1 + i \lambda_L^2 \\
\lambda_R^1 + i \lambda_R^2
\end{pmatrix} 
=
\begin{pmatrix}
\psi_L \\
\psi_R
\end{pmatrix} .
\end{align}
Then, we can unfold the $c = 1$ Dirac fermion with the above boundary condition into a non-chiral $c = \frac{1}{2}$ Majorana fermion theory with a single Majorana field $\lambda = (\lambda_L\ \lambda_R)^T$ on the full line $-\infty < x < \infty$.
The unfolded fields are defined by
\begin{align}
\begin{cases}
\psi_R(t,x) = \lambda_R(t,x) + i \lambda_L(t,-x), \\
\psi_L(t,x) = \lambda_R(t,-x) + i \lambda_L(t,x) . \notag
\end{cases}
\end{align}
In this unfolded description, we obtain a single Majorana fermion theory,
\begin{align}
S = \int _{-\infty}^\infty dt\int_{-\infty}^{\infty} dx
\Big( i \bar{\lambda} \gamma^{\mu}\partial_{\mu} \lambda - i m \bar{\lambda}\lambda-  \text{sgn}(x)i\Delta\bar{\lambda} \lambda   \Big) ,
\end{align}
with the interface condition
\begin{align}
&\lambda_L(t, -0 ) = - \lambda_L(t, +0), \notag \
&\lambda_R(t, -0 ) = \lambda_R(t, +0).
\end{align}
This boundary condition is realized by inserting the timelike chiral fermion-parity defect $(-1)^{F_L}$ at $x = 0$.

On the other hand, in the staggered fermion lattice regularization, the one-site translation of the Majorana fermions flows in the continuum limit to the chiral fermion parity $(-1)^{F_L}$ \cite{Seiberg:2023cdc}. This idea is illustrated in Fig.~\ref{fig:low-energy} (d,e).
Combining these two facts, the lattice realization of the Emery--Kivelson boundary condition is naturally identified with the one-site translation defect in the staggered Majorana fermion model. This identification is not an exact microscopic equality on the lattice; it is an emergent low-energy statement. This is why we call the phenomenon emergent Andreev reflection.

\paragraph{Relation to interfaces between CFTs and SPT phases.}
We illustrate here how the lattice EK model, Eq.~\eqref{eq:EK_complex_fermion}, naturally arises as an interface between a gapless chain and an SPT phase.
Strictly speaking, on the lattice, we consider an interface between a gapped SPT segment and a gapless Majorana chain, whose low-energy limit is described by a CFT.
An explicit Hamiltonian description is provided in the Supplemental Material.
Starting from the original EK boundary, Fig.~\eqref{fig:EKandSPT} (a), we can add and remove Majorana pairs to form a Kitaev chain, shown in Fig.~\eqref{fig:EKandSPT} (b). In the limit that the SPT side is shrunk to just its boundary, only the interface Majorana mode remains coupled to the gapless chains. This way, the boundary Majorana mode appears naturally in the EK construction.

\begin{figure}[bt]
    \centering
    \includegraphics[width=\linewidth]{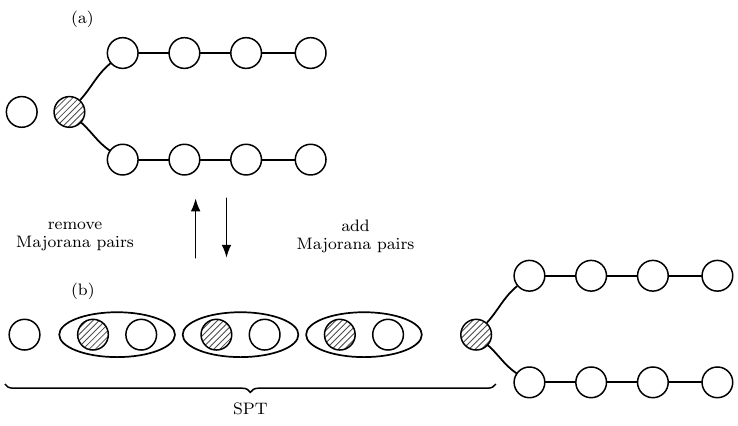}
    \caption{
    Relation between the Emery--Kivelson boundary and an interface between a Kitaev-chain SPT segment and gapless Majorana chains.
(a) The lattice EK boundary, Eq.~\eqref{eq:EK_complex_fermion}.
(b) An interface between a Kitaev-chain SPT segment and gapless Majorana chains.
The lattice EK boundary can be viewed as the minimal boundary limit of this SPT-CFT interface.
    }
    \label{fig:EKandSPT}
\end{figure}

\paragraph{Two-channel Kondo problem and monopole interpretation.}
We now return to the problems that motivated the construction:
the two-channel Kondo boundary and Maldacena--Ludwig monopole scattering.
We first recall that the charge conjugation for the $U(1)=SO(2)\subset O(2)$ symmetry is
represented by the sign flip
 \begin{align}
 C =
\begin{pmatrix}
1 & 0 \\
0 & -1
\end{pmatrix}
\in O(2).
\end{align}
Note that this transformation is not an element of $SO(2)$.
For multiple flavors, the symmetry is extended to $SO(2N_f)$. The natural
generalization of the above charge-conjugation transformation is
\begin{align}
C_{N_f} = \mathrm{diag}(+1,+1,\cdots,+1,-1)\in O(2N_f).
\end{align}
This transformation exchanges the two chiral spinor representations
$\bm{2}_+^{N_f-1}$ and $\bm{2}_-^{N_f-1}$, because it flips the chirality
operator as
\begin{align}
&\Gamma = (-i)^{N_f}\Gamma_1\Gamma_2\cdots \Gamma_{2N_f}
 \notag \\
 \qquad & \longrightarrow
(-i)^{N_f}\Gamma_1\Gamma_2\cdots (-\Gamma_{2N_f})
= -\Gamma .
\end{align}
Here $\Gamma_i, i = 1 \cdots ,2N_f$ are the gamma matrices for $SO(2N_f)$.
$C_{N_f}$ is again not an element of $SO(2N_f)$ and acts as an outer automorphism of $SO(2N_f)$.
For $N_f=4$, this gives the boundary condition that flips the spinor representation $\bm{2}_+^{3} = \bm{8}_s $ and $\bm{2}_-^{3} = \bm{8}_c$ obtained after the triality
transformation \cite{Maldacena_1997}.

In the continuum theory, the boundaries of the two-channel Kondo model and the four-flavor Callan-Rubakov scattering problem flow to a conformal boundary condition with Affleck-Ludwig boundary degeneracy $g_{\mathrm{AL}}=\sqrt{2}$ \cite{Affleck_Ludwig_1991, Affleck_Ludwig_1992, Maldacena_1997}. Equivalently, the boundary carries a nontrivial entropy $S_{\mathrm{bdy}}=\ln \sqrt{2}$. In our lattice model, this factor is visible microscopically: it is the contribution of the decoupled Majorana zero mode at the boundary. A single Majorana is half of a complex fermion, and therefore carries the square-root Hilbert-space weight $\sqrt{2}$.

\textcolor{black}{Under triality, the Maldacena--Ludwig wall maps to trivial boundary conditions for three flavors and a charge-flip boundary condition for the remaining flavor. On the lattice, this can be realized using four flavors of complex fermions, with free boundary conditions for three flavors and an EK boundary condition introduced in this Letter for the fourth. In the original variables, however, the free Majorana zero mode is distributed across all four flavors, making it harder to identify, as in the field-theoretic description of monopole scattering}~\cite{Triality_preparation}. 

\paragraph{Conclusion and Outlook.}
In this Letter, we found that the Andreev reflection, a charge-flip boundary condition for incoming and outgoing chiral fermions, is realized on the lattice as a hidden one-site translation of Majorana operators. This translation can be implemented by folding the transverse-field Ising model at the Kramers--Wannier duality defect. In the low-energy limit, this defect acts as the chiral fermion-parity defect $(-1)^{F_L}$, flipping the sign of only one of the two chiral Majorana modes. After recombining the two modes into a complex fermion in the folded geometry, this chiral sign flip becomes precisely the Andreev-like charge-flip boundary condition. We also showed how this boundary can be viewed as an SPT-CFT interface.

A promising future direction is to generalize the present Majorana construction beyond the two-channel case.
It is well-known that the Kondo IR fixed point BCFT has an impurity degeneracy which coincides with the quantum dimension of the spin-$\frac12$ simple object of $su(2)_k$. $k=2$ is precisely the Ising category with $d_\sigma = \sqrt{2}$, while $k = 3$ is the Fibonacci category, which has no $\mathbb{Z}_n$ structure. A similar lattice realization with `a decoupled Fibonacci anyon' would then provide the expected boundary degeneracy $g = d_\tau$.

Another natural generalization is to study the two-dimensional effective description of the monopole-fermion system for a general value of Dirac fermions $N_f$ \cite{Ohmori:unpublished,Tachikawa:2026cxd,Wei:2026fsn}.
This is tantamount to studying $SO(2N_f)_1 \supset SU(2)_k \otimes SU(k)_2$, where $2k$ is identified with the number of Majoranas.
$SO(2N_f)_1$ is the UV-free fermion algebra whose WZW model has a natural duality defect, namely that of its $\mathbb{Z}_{N_f}$ parafermion coset with dimension $\sqrt{N_f}$. 
This duality defect is a natural candidate as it lives in the UV and has a concrete lattice realization via parafermions, namely as something KW-like from the TY$(\mathbb{Z}_{N_f})$ fusion category \cite{Tambara_Yamagami_1998}.

\textcolor{black}{A final direction is to explore the connection to the superconducting proximity effect proposed by Fu and Kane~\cite{PhysRevLett.100.096407}. The connection is natural because the Majorana of the EK boundary can be viewed as an edge mode of the Kitaev chain in its $p$-wave superconducting phase. By placing the EK boundary on both sides of a finite interval, this mimics a normal metal region sandwiched between two superconductors. In our construction, this setup is equivalent to the transverse-field Ising model with periodic boundary conditions. The chiral Majorana modes can then undergo sequences of Andreev reflections, forming an Andreev bound state.}

\paragraph{Acknowledgements.}
A.U. and B.D. thank Frank Verstraete for introducing the Kondo problem and for insightful discussions. A.U. was supported by FWO Junior Postdoctoral Fellowship (grant No.\ 12AHS26N). B.D. was supported by BOF-GOA (Grant No.\ BOF23/GOA/021).
T.N. was supported by MEXT KAKENHI Grant No.~23K13094 and 24H00944 and by JST PRESTO Grant No.~JPMJPR2359.
M.W. was supported by Grant-in-Aid for Early-Career Scientists No.~25K17387, and by a MEXT KAKENHI Grant No.~24H00957.
All simulations were done with the open-source packages TensorKit.jl \cite{tensorkit} and MPSKit.jl \cite{mpskit}.
\bibliography{apssamp}

\clearpage
\onecolumngrid
\begin{center}
\textbf{Supplemental Material}
\end{center}
\vspace{0.2cm}
\twocolumngrid
\textcolor{black}{
\section{Spin representation of $H_{EK}$}
After the particle-hole transformation $\chi^2_{2j}\rightarrow -\chi_{2j}^2,$ $H_{EK}$ becomes 
\begin{align}
    H_{EK} =& i\sum_{\,j \, \mathrm{even}}^{N-1}\left(g\chi^1_{j+1}\chi^1_{j} - \chi^2_{j}\chi^2_{j+1} \right) \nonumber \\
    &+i\sum_{j > 1\, \mathrm{odd}}^{N-1}\left(\chi^1_{j+1}\chi^1_{j} - g\chi^2_{j}\chi^2_{j+1} \right) \nonumber \\
    &+i\chi^1_2\beta_\mathrm{act} - ig\beta_\mathrm{act}\chi^2_2,
\end{align}
We use the same Jordan-Wigner transformation introduced in the main text:
$$\chi_{2j-1} = \left(\prod_{k=1}^{j-1} Z_k\right)X_j,\, \chi_{2j} = \left(\prod_{k=1}^{j-1} Z_k\right)Y_j.$$ 
As shown in Fig. 1 $(c)$, however, the pairing of Majorana fermions is altered: the Majorana operators in the first and second chains, respectively, correspond to
$$\chi^1_j = \chi_{2j-1},\qquad \chi_j^2 = \chi_{2j}.$$
As a result, the neighboring terms are no longer the Ising terms but
\begin{align}
    i\chi^1_j\chi^1_{j+1} &= i\chi_{2j-1}\chi_{2j+1},\nonumber\\
    &= i (X_{j}\otimes I_{j+1})(Z_j\otimes X_{j+1}),\nonumber\\
    &= Y_jX_{j+1},
\end{align}
and 
\begin{align}
    i\chi^2_j\chi^2_{j+1} &= i\chi_{2j}\chi_{2j+2},\nonumber\\
    &= i (Y_{j}\otimes I_{j+1})(Z_j\otimes Y_{j+1}),\nonumber\\
    &= -X_jY_{j+1},
\end{align}
where the $Z$-string is cancelled up to site $j-1$ in the second line. 
Finally, the boundary term can be found by identifying 
$$\beta_{\mathrm{act}} = \chi_2.$$
This allows us to write down the boundary Majorana interaction as 
\begin{align}
    i\chi_2^1\beta_{\mathrm{act}}&-i g\beta_{\mathrm{act}}\chi^2_2 = i\chi_3\chi_2-ig\chi_2\chi_4\nonumber\\
    &=-i(Y_1\otimes I_2)(Z_1\otimes X_2) -ig (Y_1\otimes I_2)(Z_1\otimes Y_2)\nonumber\\
    &=X_1(X_2+gY_2).
\end{align}
Combining all terms together, the Emery--Kivelson Hamiltonian is derived: 
\begin{align}\label{eq:EKmodel}
    H_{\mathrm{EK}} =& \sum_{j \, \mathrm{even}}^{N-1} (X_{j} Y_{j+1}- gY_jX_{j+1})  \nonumber \\
    &+\sum_{j > 1\, \mathrm{odd}}^{N-1} (gX_{j} Y_{j+1}- Y_jX_{j+1}) \nonumber \\
    &+ X_1(X_2+g\,Y_2).
\end{align}
}
\section{Relation to interfaces between CFTs and SPT phases.}

Here, we elucidate the CFT-SPT interface through an explicit Hamiltonian construction.
Consider the following Majorana Hamiltonian which describes Fig.~\eqref{fig:EKandSPT} (b):
\begin{align}
    H _{\text{SPTedge}}
    & =  \sum_{i = -K}^0 ( -i \beta_i \alpha_{i+1} )+ i \beta_1 \chi_2^1 - i \beta_1 \chi_2^2 \notag \\
     & \qquad + \sum_{i = 2}^{N-1} (i \chi_{i}^1 \chi_{i+1}^1 + i \chi_{i}^2 \chi_{i+1}^2). \label{eq:SPTCFT}
\end{align}
Here $\alpha_i$ and $\beta_i$, with $i=-K,\cdots,0,1$, are Majorana fermions forming the SPT segment.
The first term describes a gapped Kitaev-chain segment in the topological phase.
It leaves an unpaired Majorana mode at the interface, represented here by $\beta_1$, which couples to the two critical Majorana chains through the second and third terms.
The last term is the gapless Majorana chain on the CFT side of the interface.

When $K=-1$, the SPT segment shrinks to its boundary Majorana mode.
In this limit, the Hamiltonian \eqref{eq:SPTCFT} reduces precisely to the lattice Emery--Kivelson model (6) (main text), up to the same convention for labeling Majorana operators used above.
Thus, the lattice Emery--Kivelson boundary can be viewed as the minimal interface between a Kitaev-chain SPT segment and gapless Majorana chains.

This interpretation becomes even more transparent after introducing complex fermions.
We define
\begin{align}
    c_j =
    \begin{cases} \displaystyle\displaystyle \frac{1}{2}e^{i\frac{ \pi}{4}} (\beta_j + i \alpha_j) \quad  &- K \le j \le 1\\[10pt]
     \displaystyle\frac{1}{2} (\chi_j^1 + i \chi_j^2) \quad  &2 \le j \le N
    \end{cases}
    \end{align}
    In terms of these complex fermions, the Hamiltonian can be written in the Bogoliubov--de Gennes form
    \begin{align}
    &H _{\text{SPTedge}} \notag \\
    &= \sum_{j = - K}^{N-1} (t _jc_j^{\dagger} c_{j+1} + t_j^* c_{j+1}^{\dagger} c_j + \Delta_j c_j c_{j+1} + \Delta_j^*  c_{j+1}^{\dagger}c_{j}^{\dagger} ).
\end{align}
with
\begin{align}
    \Delta_j = 
    \begin{cases}
        0 \qquad &j \ge 2 \\
        -1 \qquad &j \le 1 
    \end{cases}, \quad 
    t_j = 
    \begin{cases}
        i \qquad &j \ge 1 \\
        -1 \qquad &j \le 0
    \end{cases}.
\end{align}
\begin{figure}[bt]
    \centering
    \includegraphics[width=\linewidth]{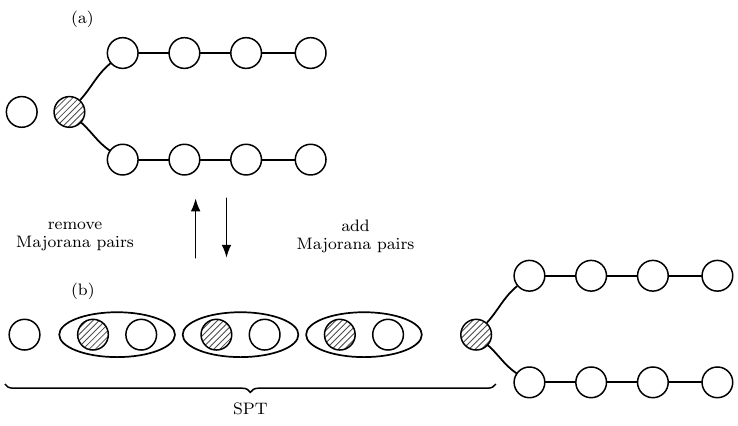}
    \caption{
    Relation between the Emery--Kivelson boundary and an interface between a Kitaev-chain SPT segment and gapless Majorana chains.
(a) The lattice EK boundary, Eq.~(6) (main text).
(b) An interface between a Kitaev-chain SPT segment and gapless Majorana chains.
The lattice EK boundary can be viewed as the minimal boundary limit of this SPT-CFT interface.
    }
    \label{fig:EKandSPT}
\end{figure}
In this form, the physical meaning of the construction is clear.
For $j\le 1$, the pairing amplitude $\Delta_j$ is nonzero, so the left segment is a superconducting, or equivalently Kitaev-chain SPT phase.
For $j\ge 2$, the pairing amplitude vanishes and the system is a normal gapless fermion chain.
Therefore, the lattice Emery--Kivelson model is the minimal boundary limit of an interface between a topological superconducting SPT phase and a gapless normal chain.
The emergent Andreev reflection discussed above is then naturally interpreted as the scattering process at this SPT-CFT interface.

\section{Translation defects and axial $U(1)_A$-symmetric boundaries.}
\begin{figure}[t]
    \centering
    \includegraphics[width=\linewidth]{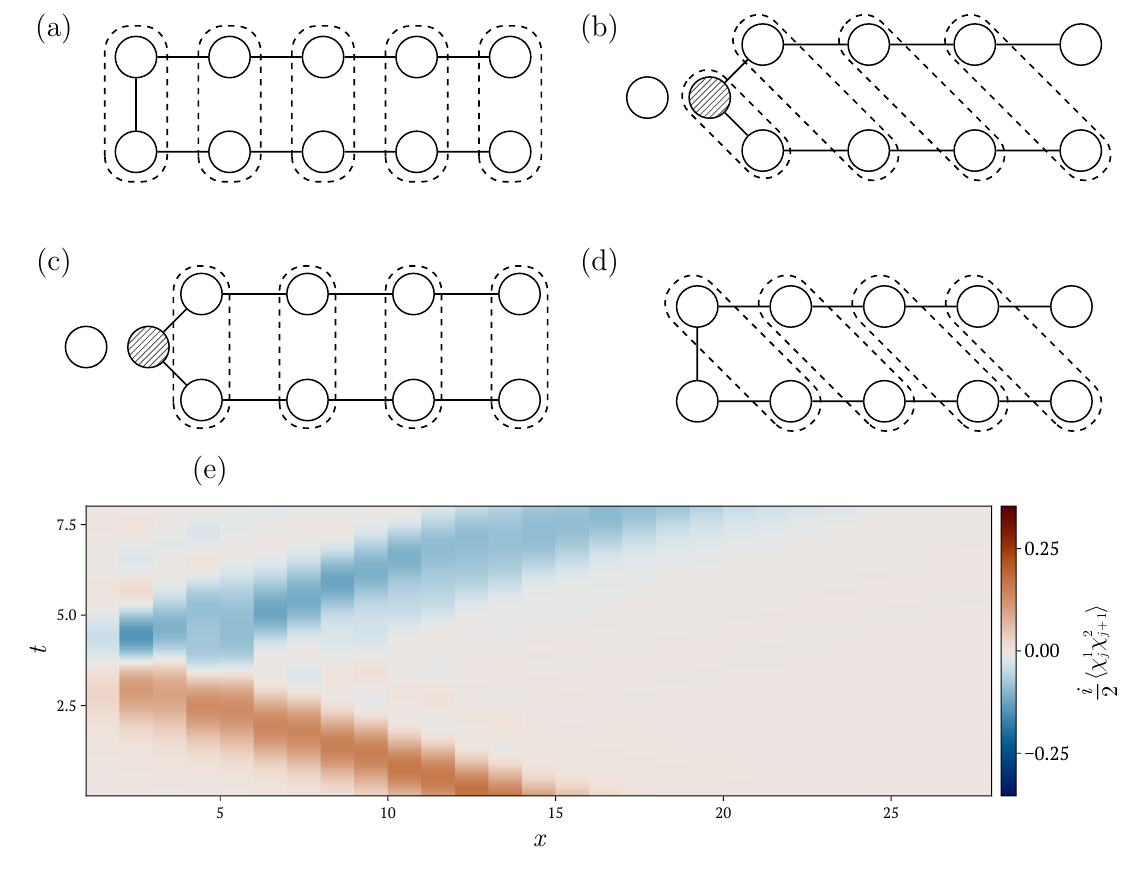}
    \caption{Emery--Kiverson boundary condition as axial $U(1)_A$-symmetric boundary.
    The dashed lines represent the charge density operators.
    (a) A manifestly vector $U(1)_V$-symmetric model \eqref{eq:sfhiftedEK}.
    Dashed lines represent vector charge densities $q_i^V = c^{\dagger}_i c_i -1/2 = \frac{i}{2}\chi_i^1 \chi_i^2$.
    (b) Emery--Kivelson boundary and axial charge density $q_{i+\frac{1}{2}}^A = \frac{i}{2} \chi_{i}^1 \chi_{i+1}^2$.
    Translation of the first Majorana fermions $\chi_{2j}^1 \to \chi_{2j+1}^1$ and renaming $\beta_{\rm act} \to \chi_2^1$ \eqref{eq:shiftVA} is understood as the change of the shape of the chain to that in (a).
    (c) Vector $U(1)_V$ charge density $q_i^V$ in the Emery--Kivelson boundary.
    $U(1)_V$ is not conserved but its charge is flipped.
    (d) Axial $U(1)_A$ charge density $q_{i+\frac{1}{2}}^A$ in the Hamiltonian \eqref{eq:sfhiftedEK}.
    This setup is again mapped to the lattice Emery--Kivelson model by the inverse of \eqref{eq:shiftVA}.
    Correspondingly, the axial $U(1)_A$ charge should be flipped at the boundary.
    (e) Simulation of the $U(1)_A$ charge flip in the Hamiltonian \eqref{eq:sfhiftedEK}.
    }
    \label{fig:placeholder}
\end{figure}
The same one-site Majorana translation also explains which continuous symmetry survives at the charge-flip boundary. In the folded complex-fermion language, the Emery--Kivelson boundary does not preserve the ordinary vector charge $U(1)_V$; the reflected excitation has the opposite charge. However, the translation defect maps the vector charge density to a staggered, or axial, charge density. Thus, the charge-flip boundary should not be viewed simply as a charge-nonconserving boundary, but rather as a boundary which preserves the dual, axial $U(1)_A$ symmetry.

This observation is significant in view of the general relation between anomalies and boundaries: anomalous symmetries obstruct symmetric boundary conditions \cite{Numasawa:2017crf,Han:2017hdv,Thorngren:2020yht,Hellerman:2021fla}, whereas non-anomalous symmetries are expected to allow them. What is nontrivial here is that this expectation is realized microscopically on the lattice: the duality/translation defect converts the ordinary vector $U(1)_V$ into an axial $U(1)_A$, and the boundary Majorana mode completes the conserved axial charge.

Indeed, at the gapless point $g=1$ the Hamiltonian (6) (main text) is invariant under the axial symmetry  \cite{Thacker:1994ns,Horvath:1998gq}
\begin{align}
    Q_A = \frac{i}{2} \beta_{\mathrm{act}}\chi_{2}^2 + \frac{i}{2}\sum_{j = 2}^{N-1} \chi_{j}^1 \chi_{j+1}^2. \label{eq:axialcharge}
\end{align}

This contrasts with the vector $U(1)_V$ symmetry associated with the number of particles
\begin{align}
    Q_V = \sum_{j = 2}^{N-1}\Big( c^{\dagger}_j c_j - \frac{1}{2}\Big) = \frac{i}{2}\sum_{j = 2}^{N-1} \chi_{j}^1 \chi_{j}^2.
\end{align}

The axial $U(1)_A$ and vector $U(1)_V$ are related by the one-site Majorana translation on the second Majorana fermion $\chi_{j}^2$ \cite{Chatterjee:2024gje}\footnote{ Briefly speaking, we do not have translation with a boundary.
The free boundary condition at $j = N-1$ also violates the axial symmetry.
Those can be avoided by imposing the lattice Emery--Kivelson boundary condition at $j=N$ rather than free boundary conditions.}.
Note that we also have to include the impurity mode $\beta_{\rm act}$ to write the axial charge preserved by the lattice Emery--Kivelson boundary.

In fact, translating the first Majorana fermions 
\begin{align}
    \chi_{2j}^1 \to \chi_{2j+1}^1, \quad \text{and relabeling  }  \quad \beta_{\rm act} \to \chi_2^1,  \label{eq:shiftVA}
\end{align}
the lattice Emery--Kivelson model (6) (main text) is mapped to the Hamiltonian
\begin{align}
    H = -c^{\dagger}_2 c_2 + i\sum_{j=2}^{N-1}  (c_j^{\dagger}c_{j+1} - c_{j+1}^{\dagger}c_j), \label{eq:sfhiftedEK}
\end{align}
which is manifestly invariant under the ordinary vector symmetry $U(1)_V$.
This mapping is illustrated in Fig.~\ref{fig:placeholder} (a-d).
Under these translations the role of $U(1)_V$ and $U(1)_A$ is exchanged.
We also numerically confirm that in \eqref{eq:sfhiftedEK} the axial charge is flipped, as demonstrated in Fig.~\ref{fig:placeholder} (e).
\end{document}